\documentclass[reprint,superscriptaddress]{revtex4-1}
\usepackage{mathtools}
\usepackage{amsmath,amssymb}
\usepackage{color}
\usepackage{braket}
\usepackage{graphicx}
\usepackage{epstopdf}
\usepackage{float}
\usepackage{nccmath}
\usepackage{times}
\usepackage[normalem]{ulem} 
\usepackage{textgreek}
\usepackage{colortbl}
\usepackage{multirow}
\usepackage{hhline}
\let\vec\mathbf
\usepackage{soul}
\usepackage{hyperref}
\usepackage[table]{xcolor}
\usepackage[toc,page]{appendix}

\begin{document}

\title{A topological route to engineering robust and bright supersymmetric laser arrays}

\author{Soujanya Datta}
\affiliation{Theory Division, Saha Institute of Nuclear Physics, 1/AF Bidhannagar, Kolkata 700064, India}
\affiliation{Homi Bhabha National Institute, Training School Complex, Anushaktinagar, Mumbai
400094, India}

\author{Mohammadmahdi Alizadeh}
\affiliation{Department of Physics, Michigan Technological University, Houghton, Michigan 49931, USA}

\affiliation{Department of Electrical and Computer Engineering, Saint Louis University,  Saint Louis, MO 63103, USA}

\author{Ramy El-Ganainy}
\email[]{relganainy@slu.edu}
\affiliation{Department of Physics, Michigan Technological University, Houghton, Michigan 49931, USA}
\affiliation{Department of Electrical and Computer Engineering, Saint Louis University,  Saint Louis, MO 63103, USA}

\author{Krishanu Roychowdhury}
\email[]{krishanu@pks.mpg.de}
\affiliation{Theory Division, Saha Institute of Nuclear Physics, 1/AF Bidhannagar, Kolkata 700064, India}
\affiliation{Homi Bhabha National Institute, Training School Complex, Anushaktinagar, Mumbai
400094, India}
\affiliation{Max-Planck-Institut f\"{u}r Physik komplexer Systeme, N\"{o}thnitzer Strasse 38, Dresden 01187, Germany}

\begin{abstract}
In recent years, several proposals that leverage principles from condensed matter and high-energy physics for engineering laser arrays have been put forward. The most important among these concepts are topology, which enables the construction of robust zero-mode laser devices, and supersymmetry (SUSY), which holds the potential for achieving phase locking in laser arrays. In this work, we show that the relation between supersymmetric coupled bosonic and fermionic oscillators on one side, and bipartite networks (and hence chiral symmetry) on another side can be exploited together with non-Hermitian engineering for building one- and two-dimensional laser arrays with in-phase synchronization. To demonstrate our strategy, we present a concrete design starting from the celebrated Su-Schrieffer-Heeger (SSH) model to arrive at a SUSY laser structure that enjoys two key advantages over those reported in previous works. Firstly, the design presented here features a near-uniform geometry for both the laser array and supersymmetric reservoir (i.e. the widths and distances between the cavity arrays are almost the same). Secondly, the uniform field distribution in the presented structure leads to a far-field intensity that scales as $N^2$ where $N$ is the number of lasing elements. Taken together, these two features can enable the implementation of higher-power laser arrays that are easy to fabricate, and hence provide a roadmap for pushing the frontier of SUSY laser arrays beyond the proof-of-concept phase.  
\end{abstract}

\maketitle

\section{Introduction}

Over the past decade, approaches based on unconventional symmetry considerations have been devised for controlling light-matter interaction. These include non-Hermitian symmetries \cite{Ganainy2018NP, Ozdemir2019NM, Miri2019S, el2019dawn}, SUSY \cite{chumakov1994supersymmetry, miri2013supersymmetric, heinrich2014supersymmetric}, and topological invariants \cite{lu2014topological, ozawa2019topological, ota2020active}. Importantly, it was also recognized that the interplay between these seemingly different concepts can lead to exotic behavior with potential transformative applications. For instance, it was demonstrated that a judicious inclusion of gain and loss to a topological SSH array can still leave the zero-mode protected by particle-hole symmetry \cite{schomerus2013topologically}, and hence, allow for building topological lasers \cite{st2017lasing, zhao2018topological,parto2018edge}. In addition, it was proposed that non-Hermitian topological structures can be used to engineer highly reflective atomic mirrors \cite{wang2022non}. 

Interestingly, while non-Hermitian and topological concepts are deeply rooted in open quantum systems and condensed matter respectively, SUSY was first proposed in the parlance of high-energy physics and later found its way to quantum mechanics (see Ref.~\cite{cooper1995supersymmetry} for a comprehensive reference], and low-energy physics with applications in condensed matter \cite{sourlas1985introduction}, nonlinear dynamics \cite{leznov1989exactly}, mesoscopic physics \cite{efetov1995supersymmetry}, superconductivity \cite{kosztin1998free}, and stochastic processes \cite{feigelman1982hidden} to just mention a few examples. Very recently, it was discovered that the physics of coupled optical oscillators that respect parity-time reversal (PT) symmetry and experience time-periodic coupling can be described within the framework of SUSY quantum mechanics \cite{el2012local}. Shortly after, the notion of SUSY was mapped to the optical system described by Maxwell's equations and utilized as a means for tailoring the spectra of various photonic systems and engineering isospectral (or pseudo-isospectral) configurations in a deterministic and controllable fashion without relying on intensive numerical schemes that are computationally costly \cite{miri2013supersymmetric, heinrich2014supersymmetric}.

Among the various potential applications of SUSY in optics \cite{miri2013supersymmetry, yu2015bloch, yu2017controlling, viedma2022high, ezawa2023supersymmetric, liu2024perfect} SUSY laser arrays \cite{Ganainy2015PRA} have particularly received considerable attention, mainly due to their potential to solve one of the long-standing problems in laser engineering, namely that of building single-mode large-area laser systems with high power emission. In general, large-area lasers support several optical modes. When coupled with nonlinear effects, this multimode nature of large-area lasers tends to degrade the quality of their emitted radiation \cite{marciante1996nonlinear}. To solve this problem, it was proposed to use laser arrays. However, for these arrays to be useful, they must all emit in phase, {\it i.e.}, and they must be mode-locked to the in-phase supermode. As it turned out, however, achieving this is not an easy task \cite{winful1988stability, ohtsubo2017instability, kapon1984supermode}, and while several techniques have been proposed, each comes with its pitfalls. To overcome this problem, strategies have been conceptualized among which photonic crystal arrangements are particularly useful in synchronizing surface-emitting laser arrays \cite{kodigala2017lasing,contractor2022scalable,luan2023reconfigurable}. For edge-emitting laser schemes as well as for standard surface-emitting devices ({\it i.e.}, standard vertical cavity emitting lasers or VCSEL) that do not rely on photonic crystal fabrication, one of the most promising techniques is to employ SUSY laser arrays which were first proposed theoretically \cite{Ganainy2015PRA} and later implemented for 1D systems using different platforms \cite{hokmabadi2019supersymmetric, midya2019supersymmetric}. Extensions to 2D laser arrays have also been suggested \cite{teimourpour2016non} and implemented \cite{qiao2021higher}. Additionally, laser systems based on the notion of supercharge operators have been first proposed in \cite{midya2018supercharge} and later in \cite{smirnova2019parity}.

Despite the excellent experimental performance of the reported SUSY laser arrays, both in 1D \cite{hokmabadi2019supersymmetric, midya2019supersymmetric} and 2D \cite{qiao2021higher}, these structures have two main drawbacks: (i) They cannot be easily scaled up, and (ii) The in-phase emission is not uniform. The scalability problem arises because the supersymmetric partner array is made of distinct elements, {\it i.e.}, cavities with varying resonant frequencies and coupling coefficients. To make things worse, the variation among the resonant frequencies is comparable to the coupling between the resonators which is very small compared to the laser frequency and hence presets the fabrication especially when the number of cavities is large. The non-uniform emission is simply a result of engaging a uniform lasing array with supermodes that have variable intensity across the array structure. 

\begin{figure}
 \centering
  \includegraphics[width=\columnwidth]{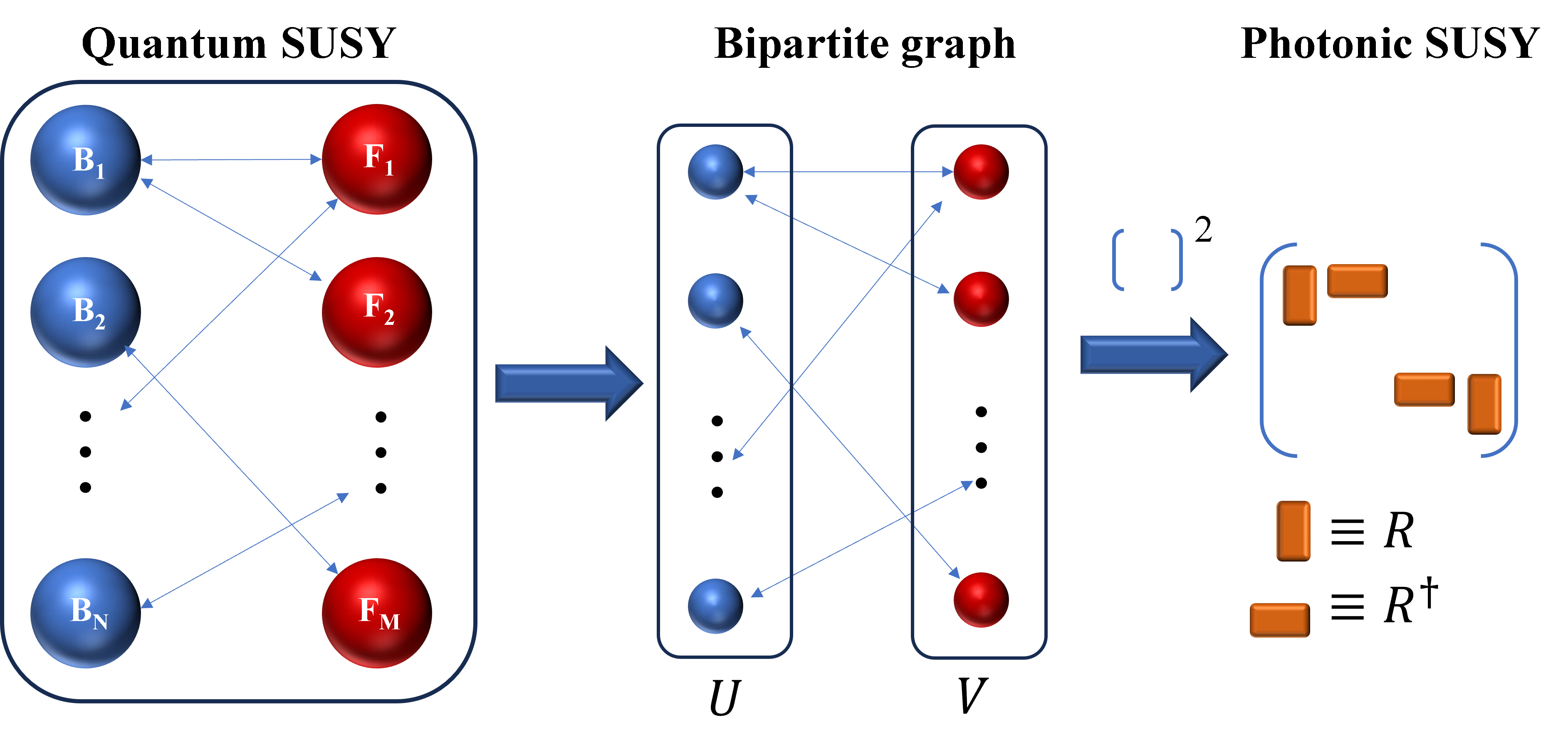}
  \caption{\textbf{From quantum SUSY to photonic SUSY.} SUSY in quantum field theory postulates that for every elementary particle, there is a corresponding partner particle that differs by one-half a spin, {\it i.e.}, for every boson, there is a corresponding fermion and vice versa. From a mathematical point of view, one can construct SUSY algebra that connects these two types of particles in any number as shown in the leftmost panel. By stripping these particles out of their spins and interpreting the connectivity matrix as a graph, we obtain a bipartite graph, which is known to exhibit chiral symmetry. By squaring the tight-binding Hamiltonian associated with this graph, we obtain a block diagonal Hamiltonian that features classical discrete SUSY as defined by matrix decomposition and operator intertwining. Note that the elements of the matrix ${\bf R}$ in this work represent dispersive coupling and hence are real. As a result, Hermitian conjugation reduces to matrix transpose.}
 \label{Fig_SUSY_Chiral}
\end{figure}

This work envisages a theoretical scheme by unifying concepts from topology and SUSY and elucidates how this can be utilized for engineering a class of supersymmetric laser arrays that can operate in the in-phase synchronous mode. Importantly, the proposed laser system features a near-uniform array design (which is important from the implementation point of view) and boasts a two-fold far-field intensity enhancement compared to the previous works. 

\section{Results}
\subsection{Formalism: From quantum to photonic SUSY via chiral symmetry}

We start by presenting a brief overview of SUSY in the context of quantum physics by considering coupled fermionic and bosonic oscillators. The defining element of the SUSY algebra is the supercharge operator \cite{cooper1995supersymmetry} that represents the coupling between the fermions and the bosons in the form 
\begin{align}\label{Eq_Supercharge}
    {\cal Q} =\sum_{\alpha,j} R_{\alpha , j} c^\dagger_\alpha  b_j + {\rm h.c},
\end{align}
where $R_{\alpha,j}$ are the elements of the matrix $\bf R$ that encodes the interaction between the bosonic and fermionic oscillators described by their respective operators $b_j$ and $c_{\alpha}$. Here we have indexed the fermions with the Greek indices and the bosons with the Latin indices. If the number of bosonic and fermionic oscillators is $M$ and $N$, respectively, then the matrix $\bf R$ has a dimension of $N \times M$. By using the algebra of the bosonic and fermionic operators, one can show (see the ``Derivation of the SUSY Hamiltonian'' subsection in Methods) that the Hamiltonian to describe the supersymmetric system including both types of operators is
\begin{align}\label{Eq_H.SUSY}
    {\cal H}_{\rm SUSY} = {\cal Q}^2 &= \frac{1}{2} \sum_{\alpha,\beta,j} R_{\alpha,j} R^*_{\beta,j} c^\dagger_\alpha c_\beta +  \frac{1}{2} \sum_{\alpha,i,j} R^*_{\alpha,i} R_{\alpha,j}  b^\dagger_i b_j \\
    &= \frac{1}{2} \sum_{\alpha,\beta} [{\bf R R^\dagger}]_{\alpha , \beta} c^\dagger_\alpha  
 c_\beta + \frac{1}{2} \sum_{i,j} [{\bf R^\dagger R}]_{i,j} b^\dagger_i  b_j\,, \nonumber
\end{align}
(for the rest of our discussion, the overall factor of $1/2$ can be safely discarded). The eigenvalues of the fermionic and bosonic terms in the Hamiltonian ${\cal H}_{\rm SUSY}$ coincide except for potential zero modes; one of the matrices (${ \bf R R^\dagger}$ or ${\bf R^\dagger R}$) may exhibit an excess of $|N-M|$ zero-energy eigenvalues in its tight-binding spectrum. 

\begin{figure}
 \centering
  \includegraphics[width=\columnwidth]{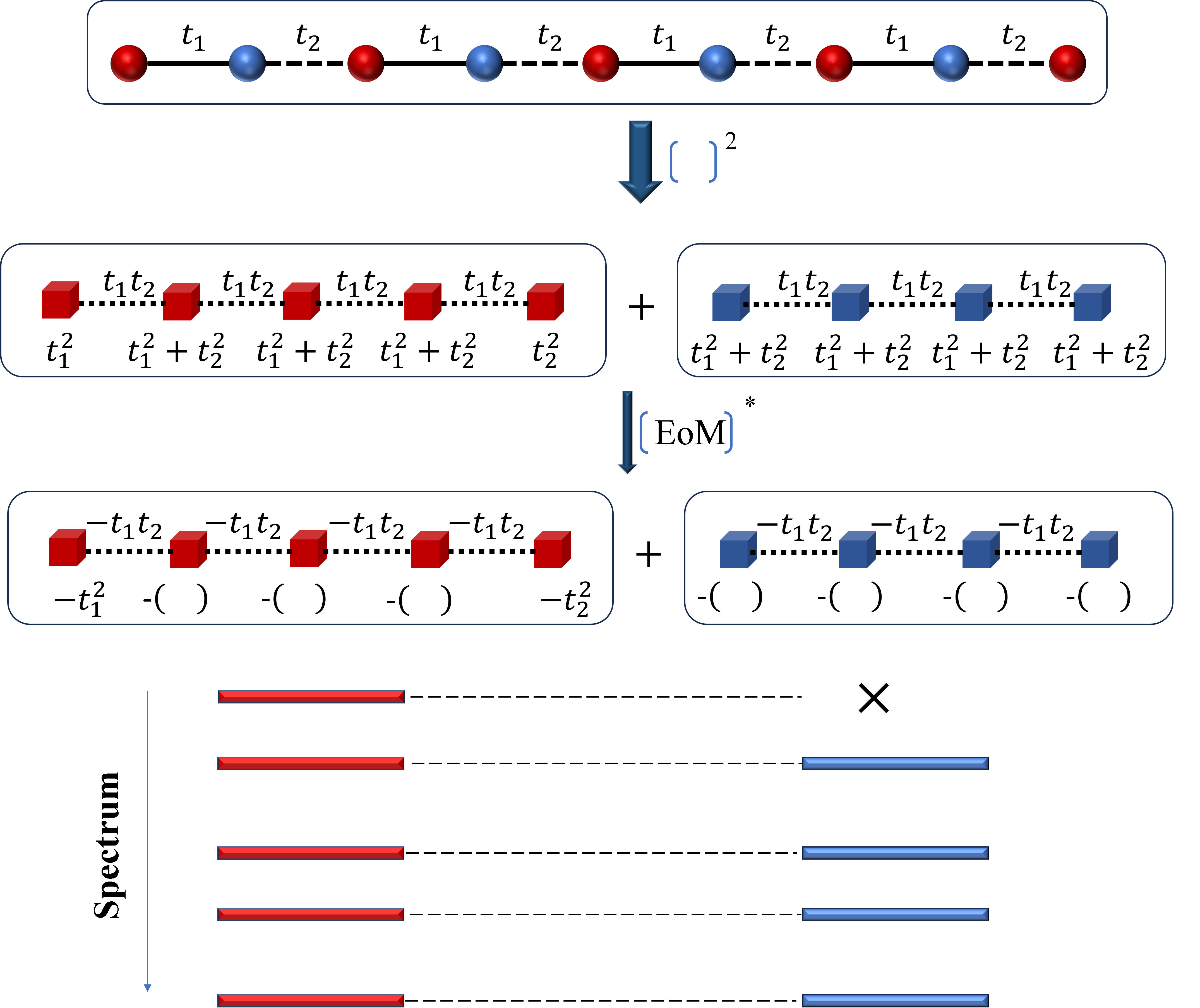}
  \caption{\textbf{SUSY structures from topological SSH arrays.} Taking the square of an SSH array (top panel) having an odd number of elements (here nine) results in two decoupled supersymmetric arrays. To ensure that the singlet mode is in phase, one of the coupling coefficients ($t_1$ or $t_2$) must be negative, which is difficult to implement experimentally. By following the procedure described in the text, which involves taking the complex conjugate of the equation of motion (EoM), we can construct another set of arrays that share the same spectral features but have negative frequency shifts and positive couplings, which is easy to implement. In the figure $-(\ \ ) \equiv -(t_1^2+t_2^2)$. Importantly, the partner array (blue cubes) is uniform while the main array (red cubes) is almost uniform except for the edge elements. This is very different from the previous work relying on QR decomposition which results in very non-uniform partner arrays difficult to implement or scale up.}
 \label{Fig_SSH_SUSY}
\end{figure}

An important aspect of the SUSY algebra is the topological nature of the supercharge operator $\mathcal{Q}$. In particular, by using the matrix representation of the fermionic creation and annihilation operators, it can be shown that $\mathcal{Q}$ obeys chiral symmetry. It follows that the spectrum of $\mathcal{Q}$ contains eigenvalues that are symmetrically distributed around the zero energy and a finite number of protected zero energy modes. Consequently, apart from the protected zero modes, the spectrum of $\mathcal{H}_{\rm SUSY}$ is doubly degenerate. 

\begin{figure}
 \centering
  \includegraphics[width=\columnwidth]{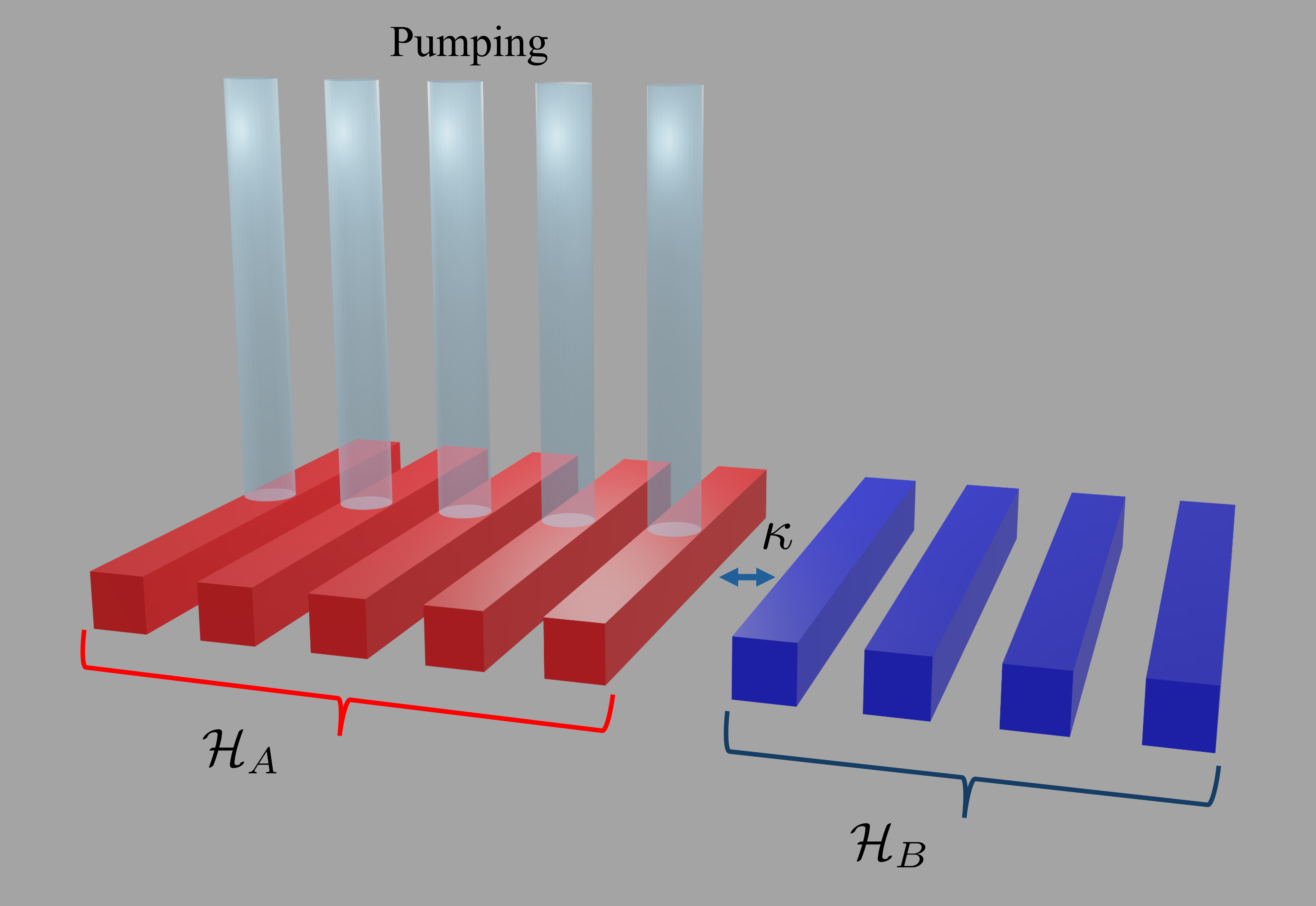}
  \caption{\textbf{Edge-emitting SUSY laser array.} An edge-emitting laser array that consists of a main and partner SUSY arrays implementing the discrete Hamiltonians $\mathcal{H}_{\rm A}$ and $\mathcal{H}_{\rm B}$, respectively. The coupling coefficients between any two lasing elements in the same array are $t$ while the edge-to-edge coupling between the two inner elements of the arrays is $\kappa$. In principle, one can also introduce a partner array on both sides of the main array to maintain the parity symmetry. Following previous experiments, here the pumping profile is assumed to illuminate only the main array.}
 \label{Fig_Edge_SUSY}
\end{figure}

To make the connection between the above picture of SUSY and classical photonic arrays, we reinterpret the supercharge operator as the weighted adjacency matrix of a bipartite graph $G(U,V,E)$ where $U$ and $V$ are the two partitions and $E$ denotes the edges connecting the graph partitions. In this picture, the fermions and the bosons are replaced by classical photonic oscillators (waveguides or cavities) that occupy the partitions $U$ and $V$ (or vice versa). Within the framework of optical coupled mode theory of cavity arrays, this classical array can be described by $i~{\rm d}\Vec{a}/{\rm d}t=\mathcal{H}_{\rm TB} ~\vec{a}$ where the vector $\vec{a}$ represent the electric field amplitude in each cavity and the classical discrete Hamiltonian is given by:   
\begin{align}\label{Eq_H_TB}
    {\cal H}_{\rm TB} =
    \begin{pmatrix}
    & {\bf R} \\
    {\bf R}^\dagger &
    \end{pmatrix}.
\end{align}
Evidently, $\mathcal{H}_{\rm TB}$ exhibits topological features dictated by its chiral symmetry. By squaring $\mathcal{H}_{\rm TB}$, the two sublattices decouple and we obtain a block diagonal matrix that consists of two positive-definite square Hermitian blocks: 
\begin{align}\label{Eq._H_TB_Square}
    {\cal H}_{\rm TB}^2 =
    \begin{pmatrix}
    {\bf R}{\bf R}^\dagger & \\
    & {\bf R}^\dagger {\bf R} 
    \end{pmatrix},
\end{align}
where the two blocks $\bf R R^{\dagger}$ and $\bf R^{\dagger} R$ represent two independent optical arrays that are isospectral except from the zero modes. A schematic of the concept discussed above is depicted in Fig. \ref{Fig_SUSY_Chiral}. We emphasize here that the SUSY between the fermionic and bosonic oscillators provides a natural framework for establishing the SUSY between classical oscillators but it does not imply that the quantum system (left panel in Fig. \ref{Fig_SUSY_Chiral} is a one-to-one mapping to the corresponding classical graph (central panel of Fig. \ref{Fig_SUSY_Chiral}. Also, we note that the above procedure eliminates only the zero modes which is different from the $QR$ decomposition scheme used in \cite{Ganainy2015PRA, hokmabadi2019supersymmetric, midya2019supersymmetric}. However, in the context of laser engineering, this is sufficient as we will see below.

Before we proceed, we remark that the relation between the spectrum of a Hamiltonian and its square has been employed before to show that crossing the non-Hermitian singularities, called exceptional points, in some parity-time (PT) symmetric systems might not be associated with phase transitions as it is usually assumed \cite{zhong2019crossing}. On the other hand, it was demonstrated that topological features can arise by taking the square root of Hamiltonian matrices \cite{arkinstall2017topological} -- a result that was later confirmed experimentally \cite{zhang2019experimental, kremer2020square}. This work in turn has inspired further activities investigating the topological aspects associated with taking the square (or higher) roots of Hamiltonian systems \cite{lin2021square, deng2022n, wu2021square}. Importantly, we also emphasize that the connection between SUSY and chiral symmetry [Eq. \ref{Eq_H_TB} and \ref{Eq._H_TB_Square}] has been highlighted before in the context of topological mechanics \cite{kane2014topological, attig2019topological}, tight-binding band structures \cite{roychowdhury2022supersymmetry}, entanglement structure in many-body states \cite{jonsson2021entanglement}, and also in the context of optics \cite{midya2018supercharge}. In particular, the work in \cite{midya2018supercharge} presented a scheme for engineering optical SUSY array partners with predetermined spectral features. 

\subsection{Application: Uniform SUSY laser arrays}

\noindent
{\bf SUSY laser structure.} We start by reviewing the generic concept of SUSY lasers proposed in Ref. \onlinecite{Ganainy2015PRA}. Starting from a laser array that consists of $N$ coupled laser elements that support $N$ linear modes, a partner array supporting $N-1$ linear modes that are isospectral with $(N-1)$ higher order modes of the main array is constructed. By introducing coupling between the main and partner array and losses to the partner array, the quality factors of all the modes except the singlet fundamental mode of the main array are spoiled. As a result, the fundamental mode of the main array exhibits a lower lasing threshold and single-mode operation can be achieved. In previous studies, matrix QR decomposition was employed to construct the SUSY partner array starting from a uniform laser array \cite{Ganainy2015PRA, teimourpour2016non}. This strategy produces a non-uniform partner array (different resonant frequencies and coupling coefficients). This feature, together with the fact that typical values of the coupling between any two elements are much smaller than the resonant frequencies (thus very precise implementation of the resonant frequencies is necessary) complicates the practical realization of this device. And despite the excellent experimental results obtained recently \cite{hokmabadi2019supersymmetric, midya2019supersymmetric}, this aforementioned drawback still presents a major hurdle towards scaling up SUSY laser arrays: Larger arrays means more non-uniformity. 

\begin{figure*}
 \centering
  \includegraphics[width=2\columnwidth]{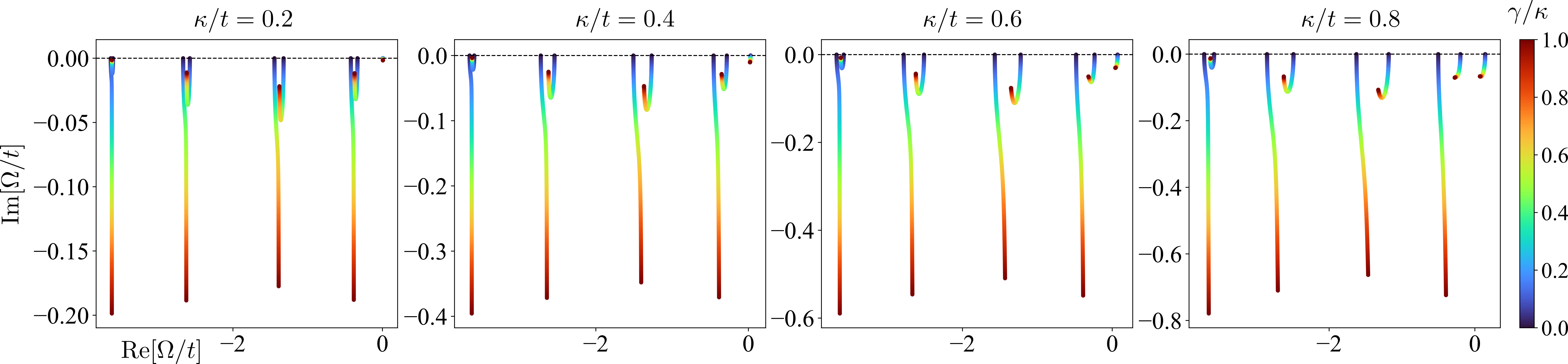}
  \caption{{\bf Complex eigenvalue trajectories of the supermodes in a lossy edge-emitting SUSY laser array.} Plotted are the complex energy spectra for the edge-emitting setup with $N=5$ at various values of the edge-to-edge coupling $\kappa/t$ and loss-to-coupling ratio $\gamma/\kappa$ taken over a range of $\gamma/\kappa\in[0,1]$. The trajectories represent how the eigenfrequencies change with the ratio $\gamma/\kappa$, the fundamental (the singlet) mode being the rightmost in all the panels. Beyond certain values of $\kappa/t$ and $\gamma/\kappa$, the fundamental mode ceases to be the first lasing mode, being taken over by one mode from the closest doublet.}
 \label{Fig_Complex_Spectra}
\end{figure*}

\begin{figure*} 
 \centering
  \includegraphics[width=2\columnwidth]{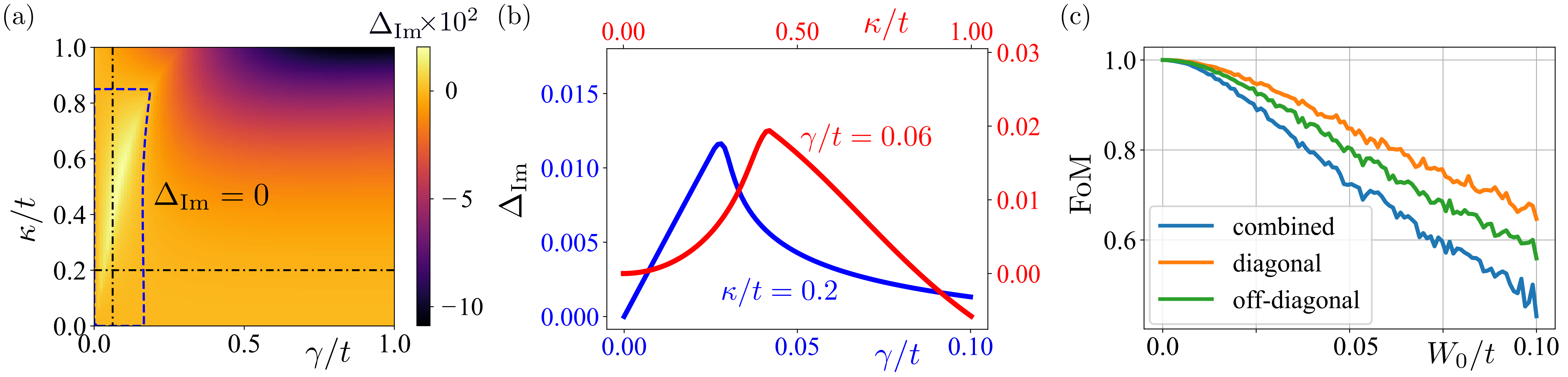}
  \caption{{\bf Detuning and disorder in the edge-emitting SUSY laser array.} (a) Shown is the 2D color map of the detuning $\Delta_{\rm}$ as a function of $\gamma/t$ and $\kappa/t$ for the edge-emitting laser with $N=5$. The dashed blue line is the contour for $\Delta_{\rm Im}=0$ with $\Delta_{\rm Im}>0$ inside that is when the singlet state is the first lasing. Outside that domain, one of the doublet states takes over to become the first lasing state. (b) The plots represent the two different cross sections marked by the two black dash-dotted lines in (a). (c) Plotted is the FoM (defined in the main text) to demonstrate how the detuning is impacted by disorders of various strengths $W_0/t$ (compared to the disorder-free case of $W_0=0$) that can be present independently in the diagonal and the off-diagonal elements of the Hamiltonian or combined. Other parameters are $\kappa/t=0.2$ and $\gamma/t=0.06$.}
 \label{Fig_Im_Detuning_2D}
\end{figure*}

To overcome the problems associated with current SUSY laser arrays, here we present a different and elegant proposal for constructing scalable SUSY laser arrays. To do so, we start by considering a topological SSH array. This is a very special bipartite graph that exhibits periodicity with two identical elements in each unit cell and different inter and intra-coupling constants as shown in Fig. \ref{Fig_SSH_SUSY}. As discussed in Ref. \onlinecite{ja2016short}, an infinite one-dimensional Hermitian SSH array exhibits topological features and can exist in two distinct phases: a topological phase with a winding number of $W=1$ (across the Brillouin zone) and a trivial phase with $W=0$. When a topological SSH array is terminated at the weak link edge, a topologically protected zero mode that is localized at that edge emerges (i.e. its frequency will be exactly the same as the bare resonant frequency of each cavity in the array, the central frequency of each resonator in the absence of any coupling). For a finite SSH array having an odd number of sites, the protection of the zero mode is guaranteed by the chiral symmetry of the underlying Hamiltonian, namely $\mathcal{H}_{\rm SSH}$ respects chiral symmetry, {\it i.e.}, $\chi^{-1} \mathcal{H}_{\rm SSH}\chi=-\mathcal{H}_{\rm SSH}$. We note however that chiral symmetry does not necessarily determine the number of zero modes. It is rather the winding number of the infinite SSH that fixes that through the bulk boundary correspondence. In the case of an SSH array with $W=1$, there is only one zero mode. Thus, if we consider a truncated SSH array made of an odd number of elements, say $(2N-1)$ then one of these modes will be pinned to have zero eigenvalues. It follows that $\mathcal{H}_{\rm SSH}^2$ will consist of two block diagonals representing two decoupled arrays, one with $N$ states including the zero mode (we call this array the main array and denote it by array A) and the other with $(N-1)$ states, which we call the partner array denoted by B. These two arrays are represented by the red and blue cubes in Fig. \ref{Fig_SSH_SUSY}. If we denote the coupling coefficients of the SSH array by $t_1$ and $t_2$, we find that both the main and partner arrays have uniform coupling profiles given by $t_1t_2$. In addition, each element of the partner array will have a frequency shift of $t_1^2+t_2^2$. Finally, all the non-edge elements of the main array will have the same frequency shift, {\it i.e.}, $t_1^2+t_2^2$ while the edge elements will have a different frequency shift given by $t_1^2$ and $t_2^2$ as shown in Fig. \ref{Fig_SSH_SUSY}. This rather very uniform structure of the SUSY arrays presented here is one of the main results of this work. From an experimental point of view, this uniformity can facilitate the experimental realization of SUSY laser arrays that consist of a relatively large number of coupled edge-emitting lasers. 

Before we proceed, we note that in the above construction, the zero mode of the main array has a non-uniform staggering eigenvector. By selecting $|t_1|=|t_2|$ the optical intensity of the zero-mode becomes uniform. However for $t_{1,2}>0$, the mode still features an out-of-phase distribution ({\it i.e.}, the sign of the field amplitude in each element is the opposite of that in the neighboring elements). This out-phase mode is not useful for laser applications since the fields will interfere destructively in the far-field. One can turn this out-of-phase mode into an in-phase mode by reversing the sign of one of the coupling coefficients. This however will make the coupling $t_1t_2$ of the SUSY array negative. Negative coupling can be implemented in very special setups such as photonic crystals for example. However, in standard edge-emitting lasers, this is not possible. To solve this problem, let us consider the coupled mode equation for a generic cavity array described by a discrete Hamiltonian $\mathcal{H}$: $i~{\rm d}\ket{a}/{\rm d}t=\mathcal{H} \ket{a}$, where, as discussed above, the diagonal elements of $\mathcal{H}$ are real positive while the off-diagonal coupling coefficients are real negative. Here, the elements of the vector $\ket{a}=\begin{bmatrix}
a_1 & a_2 & \dots & a_N \end{bmatrix}^T$ represent the field amplitudes inside each cavity.  By taking the complex conjugate of the above equation and denoting $a^*=b$, we obtain $i~{\rm d}\ket{b}/{\rm d}t=-\mathcal{H}\ket{b}$ (note that $\mathcal{H}^*=\mathcal{H}$ because all the elements of $\mathcal{H}$ are real). This system has the same spectral structure (apart from a minus sign) and thus maintains the SUSY nature of the old system. Importantly, in this equation, the coupling profile is positive and the frequency shift is negative which can be easily implemented experimentally. In the rest of this manuscript, we will focus on this latter system with negative frequency shifts and positive couplings. 

To simplify the notation, we replace $t'^2$ with $t>0$. The explicit forms of the Hamiltonians associated with the main and partner arrays (arrays A and B) are thus given by
\begin{align}\label{mainham1}
 {\cal H}_{\rm A} =
 \begin{pmatrix}
     -t   & t    & 0      & 0      & \cdots & 0      & 0      & 0      & 0 \\
     t    & -2t  & t    & 0      & \cdots & 0      & 0      & 0      & 0 \\
     0      & t    & -2t  & t    & \cdots & 0      & 0      & 0      & 0 \\
     \vdots & \vdots & \vdots & \vdots & \vdots & \vdots & \vdots & \vdots & \vdots  \\
     0      & 0      & 0      & 0      & \cdots & t    & -2t  & t    & 0 \\
     0      & 0      & 0      & 0      & \cdots & 0      & t    & -2t  & t \\
     0      & 0      & 0      & 0      & \cdots & 0      & 0      & t    & -t
 \end{pmatrix},
\end{align}
and 
\begin{align}\label{partnerham1}
 {\cal H}_{\rm B} =
 \begin{pmatrix}
     -2t   & t    & 0      & 0      & \cdots & 0      & 0      & 0      & 0 \\
     t    & -2t  & t    & 0      & \cdots & 0      & 0      & 0      & 0 \\
     0      & t    & -2t  & t    & \cdots & 0      & 0      & 0      & 0 \\
     \vdots & \vdots & \vdots & \vdots & \vdots & \vdots & \vdots & \vdots & \vdots  \\
     0      & 0      & 0      & 0      & \cdots & t    & -2t  & t    & 0 \\
     0      & 0      & 0      & 0      & \cdots & 0      & t    & -2t  & t \\
     0      & 0      & 0      & 0      & \cdots & 0      & 0      & t    & -2t
 \end{pmatrix}.
\end{align}
As discussed before, the square matrices $\mathcal{H}_{\rm A}$ and $\mathcal{H}_{\rm B}$ have sizes $N \times N$ and $(N-1) \times (N-1)$, respectively. They are supersymmetric partners featuring an unbroken SUSY, {\it i.e.}, they are isospectral except from the zero mode which, in our system, is the fundamental mode of array $\rm A$. Finally, the SUSY laser is constructed by coupling the main and partner arrays and introducing optical losses to the latter. This will spoil the quality factor of all the higher-order modes and allow a single-mode operation of the fundamental mode when the system is pumped. From an experimental point of view, the introduction of losses to the partner array can be done by depositing an absorbing metal on top of its constituent resonators or by patterning diffraction gratings to achieve radiation loss. In all cases, care must be taken to not perturb the eigenfrequency of the supermodes (imaginary and real parts of the resonant frequency are interlinked via the Kramers-Kronig relation) \cite{guo2009observation}. Alternatively, one may just apply pumping to the main array as has been done in Ref. \onlinecite{hokmabadi2019supersymmetric, midya2019supersymmetric}. In our theoretical model, the effect of uniform loss is taken into account by replacing $-2t$ by $-2t-i\gamma$ on the diagonal elements of the matrix $\mathcal{H}_{\rm B}$. Regarding the coupling between the two arrays, we note that in the case of 1D edge-emitting lasers, only edge-to-edge coupling is possible. On the other hand, in 2D VCSEL lasers, more flexible coupling profiles can be enabled. In what follows we explore both possibilities. \\

\noindent
{\bf Edge-emitting lasers.} Here we consider an edge-emitting SUSY laser array with edge-to-edge coupling between the main and partner arrays as shown in Fig. \ref{Fig_Edge_SUSY}. The coupling coefficient between the edge elements is denoted by $\kappa$. This is similar to the case studied in Ref. \onlinecite{Ganainy2015PRA} except that here the arrays are built using the strategy outlined above which produces uniform geometry instead of using the QR decomposition that produces a highly non-uniform partner array.  

\begin{figure}
 \centering
  \includegraphics[width=\linewidth]{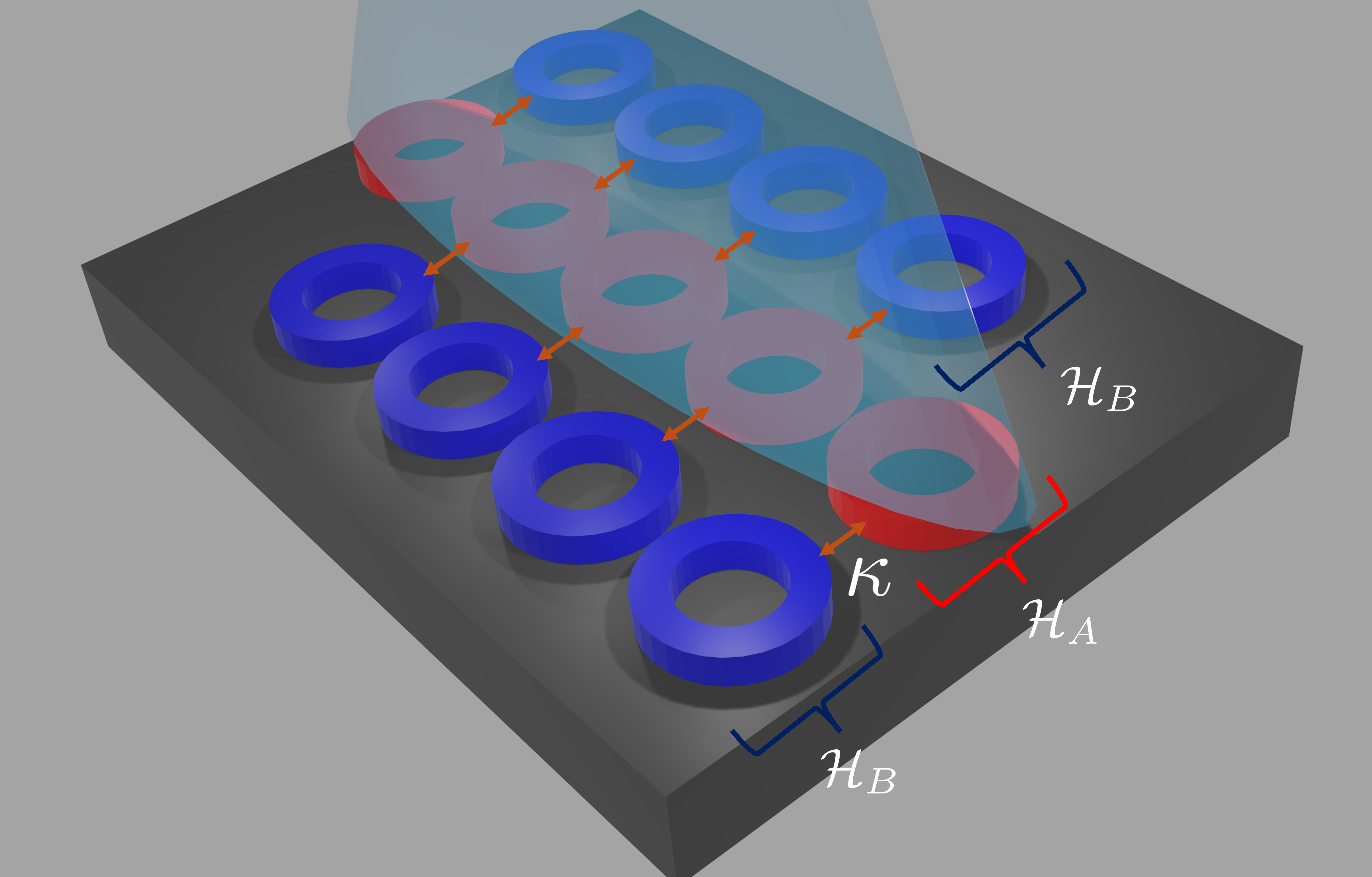}
  \caption{\textbf{Surface-emitting SUSY laser array}. A schematic of surface-emitting SUSY laser array. The 2D nature of this geometry allows for flexibility in designing the coupling between the main lasing array and the partner structure acting as a reservoir for filtering higher-order supermodes. In this particular implementation, we use two partner arrays in order to maintain the symmetry of the optical power distribution inside the laser. As before, here we assume that pumping is applied only to the lasing array.}
 \label{Fig_Surface_Laser}
\end{figure}

\begin{figure*}
 \centering
  \includegraphics[width=2\columnwidth]{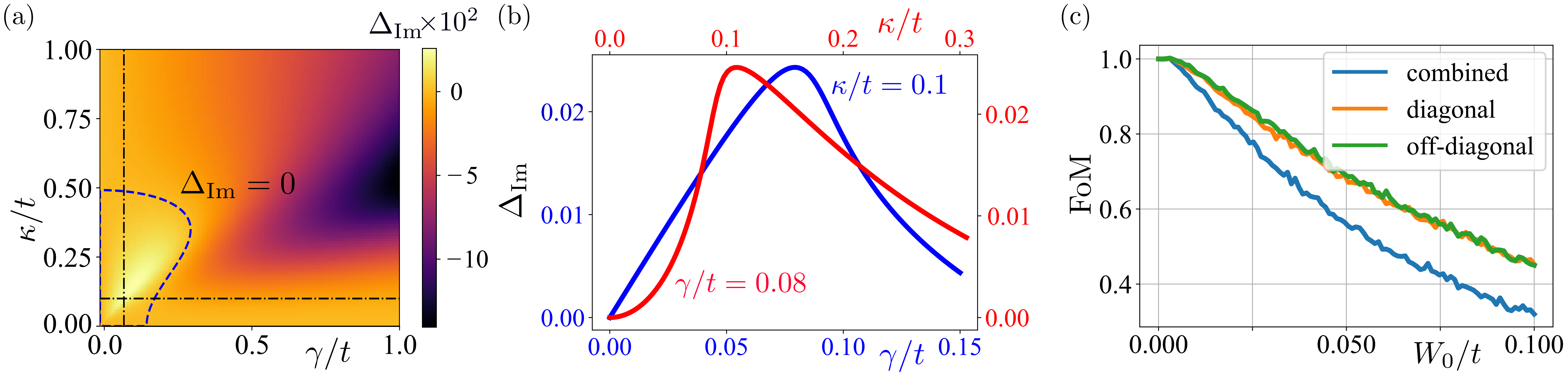}
  \caption{\textbf{Detuning and disorder in surface-emitting SUSY laser arrays.} (a) plot of $\Delta_{\text{Im}}$ as a function of $\gamma/t$ and $\kappa/t$. The dashed curved line marks the contour of $\Delta_{\text{Im}}=0$, defining the boundary of domain with $\Delta_{\text{Im}}>0$. (b) Plots of $\Delta_{\text{Im}}$ against $\gamma/t$ and $\kappa/t$ along the linear trajectories defined by the two straight dash-dotted lines in (a). Note that here, $\Delta_{\text{Im}}$ can reach higher values than in the 1D edge-emitting lasers, which indicates an even better performance for the 2D structures. (c) Device performance as a function of various possible disorder parameters, demonstrating a reasonable range of tolerance.}
 \label{Fig_Im_Detuning_2D_surface}
\end{figure*}

Fig.~\ref{Fig_Complex_Spectra} plots the complex eigenvalue trajectories of the supermodes of the entire system when $N=5$ as a function of $\gamma/\kappa$ for different values of $\kappa/t$. We note that in the first two panels, the complex resonant frequency of the fundamental mode (the singlet mode with $\text{Re}[\Omega/t]\sim 0$) remains almost intact as $\gamma$ is increased. On the other hand, in the other two panels, the eigenfrequency of the fundamental mode is shifted in response to changing $\gamma$. This effect is a result of off-resonant interaction between the modes of the main and partner arrays as has been highlighted in Ref. \onlinecite{Ganainy2015PRA}. In all cases, we observe that as $\gamma$ is increased from its zero value, the quality factors of all doublet states initially decrease ($\text{Im}[\Omega/t]$ becomes more negative). However, at some point when $\gamma$ becomes comparable to the coupling between the corresponding resonant supermodes, a complex avoided crossing takes place, and one of each doublet mode switches direction where its quality factor starts to increase again. Importantly, for the cases of $\kappa/t=0.6$ and $\kappa/t=0.8$, as $\gamma$ is increased beyond a certain limit, the singlet state becomes more lossy than one of the doublet states (the leftmost state in all the panels), {\it i.e.}, it is not the first lasing state anymore. For the cases of $\kappa/t=0.2$ and $\kappa/t=0.4$, the singlet state is the first lasing state but that same doublet state is competing for the laser action. This can be mitigated by including another lossy cavity on the other side of the array to act as a filter only for the doublet states.

To better quantify the behavior observed in Fig. \ref{Fig_Complex_Spectra}, we scrutinize the imaginary detuning $\Delta_{\rm Im}=\text{Im}[\Omega_0/t]-\text{Im} [\Omega_1/t]$ where $\Omega_0$ and $\Omega_1$ are the eigenfrequencies of the singlet state and the doublet state with nearest imaginary part ({\it i.e.}, the second lasing state), respectively. In Fig.~\ref{Fig_Im_Detuning_2D}, we plot a 2D color map of $\Delta_{\rm Im}$ as function of $\kappa/t$ and $\gamma/t$. The dashed blue line marks the contour of $\Delta_{\rm Im}=0$. Inside the region defined by this contour, $\Delta_{\rm Im}>0$ indicating that the singlet state is indeed the first lasing. On the other hand, outside that domain, one of the doublet states takes over the singlet and becomes the first lasing state. This interesting behavior has not been observed before in the previous SUSY laser work. Fig.~\ref{Fig_Im_Detuning_2D} (b) plots the two different cross-sections marked by the two black dash-dotted lines in Fig.~\ref{Fig_Im_Detuning_2D} (a). These plots confirm the behavior observed in Fig.~\ref{Fig_Complex_Spectra}, namely that $\Delta_{\rm Im}$ increases as $\gamma/t$ is increased before the trend is reversed following a complex avoided crossing. 

We have also examined the impact of disorder due to possible fabrication errors on the performance of the device. Similar to the work in Ref. \onlinecite{Ganainy2015PRA}, we define the figure of merit for the device as the ratio of the actual value of $\Delta_{\rm Im}$ in the presence of disorder to the ideal value in the absence of any perturbation, {\it i.e.}, ${\rm FoM} = \Delta_{\rm Im}^{\rm Dis}/\Delta_{\rm Im}^{\rm Ideal}$. In our simulations, we study both diagonal and off-diagonal disorders independently as well as their combined effect. In all cases, the disorder effect is considered additively, and its actual value is picked from a uniform distribution $[-W_0, W_0]$ where we varied the limit $W_0$ from 0 to $0.1t$. Finally, an averaging of $1000$ random realizations of the disordered lattice is performed for the system size $N=5$. The results are plotted in Fig.~\ref{Fig_Im_Detuning_2D} (c). The near linear degradation of the FoM is comparable to those obtained before in Ref.~\onlinecite{Ganainy2015PRA}.

At this point, it is instructive to discuss some operational considerations. As can be seen from Fig.~\ref{Fig_Im_Detuning_2D} (b), the optimal value of normalized imaginary detuning is $\Delta_{\rm Im} \sim 0.01$. Given that maximum evanescent coupling between cavities is typically in the order of THz, it follows that the maximum achievable imaginary detuning is approximately 10 GHz. For typical distributed feedback lasers with quality factors in the range $10^5-10^7$, and hence for a frequency range of $10^{2}-10^{3}$ THz, the lasing threshold can be tuned to be less than $10$ GHz (by adjusting the quality factor via grating mirrors). In that case, the above imaginary detuning will provide a large enough differentiation between the first lasing singlet mode and the competing doublet. This is consistent with the experimental results in Ref.~\onlinecite{hokmabadi2019supersymmetric, qiao2021higher}.

Finally, we also note that the uniform field distribution leads to a higher far-field maximum intensity as compared to uniform arrays with discrete sinusoidal field patterns. When considering the realistic situation of setting an upper limit on the power emitted by any individual element in either array, we find that for arrays made of five waveguides, the SUSY array of this work can provide a far-field intensity enhancement of about $\sim 1.8$ as compared to the uniform array considered in Ref.~\onlinecite{Ganainy2015PRA, hokmabadi2019supersymmetric} (see the ``Far-field intensity'' subsection and ``Array factor intensity'' subsection in Methods). In other words, while the array studied there could in principle provide $13.8$ times more intensity in the far-field than a single element laser, the SUSY array presented here in this work can provide up to $25$ times stronger intensity. And while the inclusion of the partner array will introduce some perturbations to the ideal field distribution (this has been also noted in Ref. \onlinecite{hokmabadi2019supersymmetric}, these are minimal due to the weak coupling between the main and the partner array (see the ``Power distribution inside the laser arrays'' subsection in Methods). \\

\noindent
{\bf Surface-emitting lasers.} The strategy presented in the previous section for edge-emitting devices can be also extended to surface-emitting lasers such as VCSEL arrays. In the latter case, the array is 2D which provides more flexibility for choosing the coupling topology. In Fig.~\ref{Fig_Surface_Laser}, we show one possible configuration for the array where we choose to include two partner arrays, one from each side, in order not to introduce strong asymmetry to the power distribution. As shown by the red arrows, each cavity in the partner arrays (blue rings) is coupled only to the nearest neighbor in the main, lasing array (red rings). As before, here also the loss is assumed to apply only to the partner arrays (in equal strength). Fig.~\ref{Fig_Im_Detuning_2D_surface} (a) plots the values of $\Delta_{\text{Im}}$ as a function of $\gamma/t$ and $\kappa/t$. The dashed curved line on the figure indicates the contour of $\Delta_{\text{Im}}=0$, defining the boundary of the domain with $\Delta_{\text{Im}}>0$. Panel (b) presents separate plots of the values of $\Delta_{\text{Im}}$ against $\gamma/t$ and $\kappa/t$ along the linear trajectories defined by the straight dash-dotted lines in panel (a). Note that here, $\Delta_{\text{Im}}$ can reach higher values than in the 1D edge-emitting lasers suggesting even better performance for 2D structures. Finally, panel (c) presents a study of the device performance as a function of all possible disorder parameters, demonstrating a reasonable range of tolerance. 

\section{Discussion}
In this work, we envision an angle for unifying concepts from topology and SUSY; starting from a supersymmetric system of coupled bosonic and fermionic oscillators, we establish a connection to a bipartite lattice geometry whose connectivity features chiral symmetry when imprinted on a system of classical oscillators. This picture provides a powerful tool for engineering a unique class of SUSY laser arrays. For illustration, we consider a concrete design example that takes its starting point as an SSH array to eventually manifest as a SUSY laser geometry. This array structure has two key advantages over previously reported laser arrays: (1) Both the main and partner arrays are almost uniform, {\it i.e.}, they have identical frequencies (except for only two boundary elements of the main array) and coupling coefficients across the whole arrays which is very crucial from the practical implementation perspective, and (2) The array presented here has two-fold far-field intensity scaling compared to that of Ref. \onlinecite{Ganainy2015PRA, hokmabadi2019supersymmetric}. Not only do our results present a unification for the otherwise two distinct physical concepts of topology and SUSY, but they also provide a roadmap for extending the realm of SUSY laser arrays beyond the proof-of-concept phase. Beyond its implication for laser physics and engineering, our results are generic and can be important to the physics of synchronization in various platforms including optomechanics and electronics, only to mention a couple of examples. Finally, it will be interesting to consider extending our results to complex 2D arrays. This however might not be straightforward due to the possible degeneracy associated with higher dimensional symmetries and also the long-range interactions that may arise from squaring the Hamiltonian. We plan to investigate this in future work.

\section{Methods}

\noindent
{\bf Derivation of the SUSY Hamiltonian}

\noindent
With the supercharge operator ${\cal Q}$ defined as 
\begin{align}\label{Eq_Supercharge11}
    {\cal Q} =\sum_{\alpha,j} R_{\alpha , j} c^\dagger_\alpha  b_j + R^*_{\alpha,j} b^\dagger_j c_\alpha,
\end{align}
the supersymmetric Hamiltonian reads
\begin{align}\label{Eq_H.SUSY11}
    {\cal H}_{\rm SUSY} &= {\cal Q}^2 \equiv \frac{1}{2}\{{\cal Q},{\cal Q}\}  \nonumber \\ 
    &= \frac{1}{2}\sum_{\alpha,\beta,j,k} R_{\alpha,j} R^*_{\beta,k} \{c^\dagger_\alpha b_j, b_k^\dagger c_\beta \} \nonumber \\ 
    &+ \frac{1}{2}\sum_{\alpha,\beta,j,k} R^*_{\alpha,j} R_{\beta,k} \{ c_\alpha b^\dagger_j, b_k c_\beta^\dagger \} \nonumber \\ 
    &+ \frac{1}{2}\sum_{\alpha,\beta,j,k} R_{\alpha,j} R_{\beta,k} \{ c_\alpha^\dagger b_j, b_k c_\beta^\dagger \} \nonumber \\ 
    &+ \frac{1}{2}\sum_{\alpha,\beta,j,k} R^*_{\alpha,j} R^*_{\beta,k} \{ c_\alpha b_j^\dagger, b_k^\dagger c_\beta \}.
\end{align}
We can further simplify this using the following anticommutation relations of the fermionic operators 
\begin{align}\label{commanticomm1}
  \{c_\alpha,c^\dagger_\beta\}=\delta_{\alpha,\beta}~;~ 
  \{c^\dagger_\alpha,c^\dagger_\beta\}=0~;~ 
  \{c_\alpha,c_\beta\}=0\,,  
\end{align}
and commutation relations of the bosonic operators 
\begin{align}\label{commanticomm2}
  [b_j,b^\dagger_k]=\delta_{j,k}~;~ [b_j^\dagger,b^\dagger_k]=0~;~ [b_j,b_k]=0 \,,  
\end{align}
and their mutual commutation relation
\begin{align}\label{commanticomm3}
  [c_\alpha,b_j]=[c^\dagger_\alpha,b_j]=[c_\alpha,b^\dagger_j]=[c^\dagger_\alpha,b^\dagger_j]=0 , 
\end{align}
along with the operator identity
\begin{align}
    [AB,CD] &= A[B,C]D + AC[B,D] + [A,C]BD \nonumber \\ 
            &~~~~~~~~~~~~~~~~~~~~~~~~~~~~~~~~~+ C\{A,D\}B.
\end{align}
Applying these to each of the four terms in Eq.~\ref{Eq_H.SUSY11} readily implies that the last two terms vanish while the surviving terms lead to Eq.~\ref{Eq_H.SUSY}. \\

\noindent
{\bf Far-field intensity}

\begin{figure}
 \centering
  \includegraphics[width=\columnwidth]{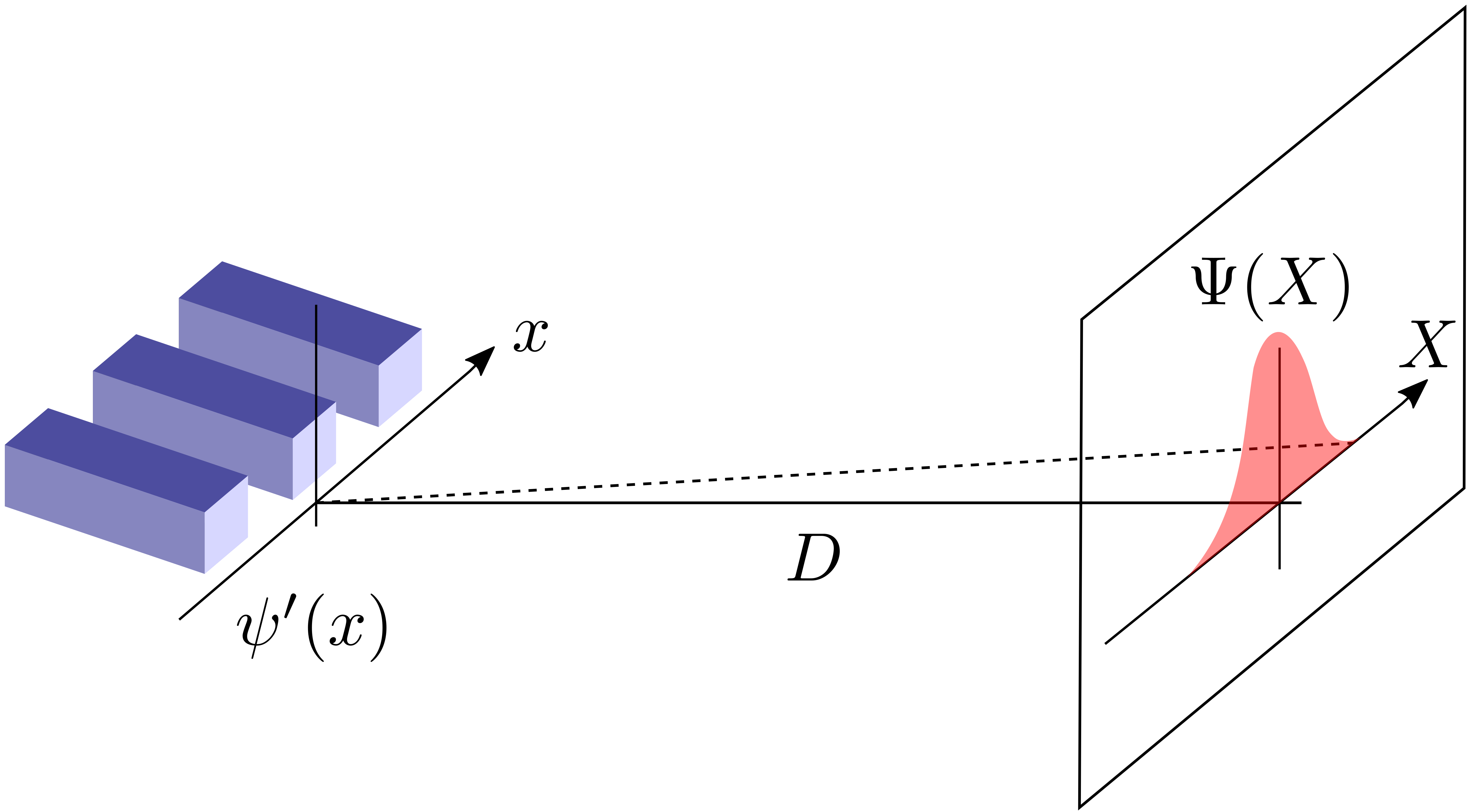}
  \caption{\textbf{Geometry for the far-field calculations.} A schematic of the geometry used for the far-field calculations.}
 \label{Fig_farfield_schematic}
\end{figure}

\noindent
Here we compare the far-field maximum intensity for the edge-emitting SUSY array considered in Fig. \ref{Fig_Edge_SUSY} and that of a uniform array ({\it i.e.}, an array with identical waveguide and identical coupling constants) such as the one studied in Ref.~\onlinecite{Ganainy2015PRA, hokmabadi2019supersymmetric} operating at the fundamental mode that has a discrete sinusoidal profile. Within the paraxial approximation, the far-field pattern can be expressed as a Fourier transform (FT) of the near-field 
\begin{align}
        \Psi(X) &= \int dx~ e^{ik(X/D)x} \psi'(x), 
\end{align}
where $k$ is the wavevector and $D$ is the distance between the near and the far-field planes. In addition, $x$ and $X$ are the near and far-field coordinates as shown schematically in Fig.  \ref{Fig_farfield_schematic}. Note that we consider a 1D FT because the array extends in 1D and thus only affects the diffraction along that direction. In our case, the near-field distribution is a weighted sum of shifted identical replicas of the waveguide mode $f(x)$ (which is assumed to be real), {\it i.e.}, $\psi'(x)=\sum_i a_i f(x-x_i)$, where $x_i$ is the lateral shift of waveguide $i$ from a reference origin point. Here we will use the normalization  $\int f(x)^2 dx=1$. To meaningfully compare the far-field response of two different arrays, we will fix the number of elements $N$ and either the total emitted power from each array or the maximum power emitted by any element of the arrays. Let us first consider the former scenario of identical total power which can be obtained from the near-field coefficients: $P_{\rm tot}=\sum_i |a_i|^2$. Without any loss of generality, we take $P_{\rm tot}=1$ in some arbitrary power units. For the laser array, the far-field can be expressed as
\begin{align}
        \Psi(X) &= \int dx~ e^{ik(X/D)x} \psi'(x) \nonumber \\
        &= \sum_i a_i \int dx~ e^{ik(X/D)x} f(x-x_i) \nonumber \\
        &= \underbrace{\int dx~ e^{ik(X/D)x} f(q)}_{F(X)} \underbrace{\sum_i a_i ~ e^{ik(X/D)x_i}}_{{\cal A}(X)}  \nonumber \\
        &\equiv F(X){\cal A}(X),
\end{align}
where $F(X)$ and $\mathcal{A}(X)$ are the aperture and the array factor, respectively. For the SUSY array considered in this work with uniform intensities across all the lasing elements, the condition $P_{\rm tot}=1$ leads to $a_j=1/\sqrt{N}$ for any $j$. Consequently, $|\mathcal{A}_1^{\rm SUSY}(X=0)|^2= \left|\sum_j a_j \right|^2 = N$ for the SUSY array shown in Fig. \ref{Fig_Edge_SUSY}.

On the other hand, for a uniform array, the normalized eigenstates are given by $a_j=\sqrt{2/(N+1)}\sin[\pi j/(N+1)]$, and hence the intensity term associated with the array factor is
\begin{align}
 |{\cal A}_1^U(X=0)|^2 = \frac{2}{N+1}{\rm cot}^2\left[\frac{\pi}{2(N+1)}\right] \equiv {\cal G}_1(N),
\end{align}
where the superscript $U$ in the above expression (see the following ``Array factor intensity'' subsection) refers to the uniform array. By using the expansion $\rm cot (x)\approx 1/x$ when $x \ll 1$, it is straightforward to show that $1< |\mathcal{A}_1^S(X=0)|^2/ |\mathcal{A}_1^U(X=0)|^2< \pi^2/8 $ with the lower and upper limits achieved at $N=1$ and $N \gg 1$, respectively. Thus the SUSY array considered here provides a (slightly) stronger far-field intensity ($\sim 1.23$ times). 

Now we consider the second scenario where we assume there is an upper bound for the power emitted by any element of the array. From an engineering perspective, this scenario is more relevant because it is the actual limiting factor in building a single-mode, single-element laser. Let us denote this maximum power by $P_o$. For the SUSY array considered here in this work, each element in the array can emit $P_o$ (implying $a_j=\sqrt{P_o}$ for all $j$). In this case, the total power is $P_{\rm tot}=N P_o$ and the maximum far-field array factor intensity is $|\mathcal{A}_2^{\rm SUSY}(X=0)|^2=N^2 P_o$. On the other hand, for the uniform array with a sinusoidal mode profile, the maximum power can be attained at the central element. For simplicity, let us assume that $N$ is odd. Then the maximum power will be emitted by the element $M=(N+1)/2$, {\it i.e.}, $a_M=\sqrt{P_o}$. But in general, $a_j=C \sqrt{2/(N+1)}\sin[\pi j/(N+1)]$ (for some constant $C$), which for $j=M$ reduces to $a_M=C \sqrt{2/(N+1)}$. Therefore, we obtain $C=\sqrt{(N+1)P_o/2}$ and finally, $a_j=\sqrt{P_o} \sin[\pi j/(N+1)]$. The maximum far-field array factor intensity is then
\begin{align}
 |{\cal A}_2^U(X=0)|^2 = P_o \ {\rm cot}^2\left[\frac{\pi}{2(N+1)}\right] \equiv {\cal G}_2(N).    
\end{align}
Hence, in his case, $1< |\mathcal{A}_1^{\rm SUSY}(X=0)|^2/ |\mathcal{A}_1^U(X=0)|^2< \pi^2/4$. Thus, the SUSY array can at most provide an enhancement of $\sim 2.47$ in the far-field maximum intensity. For a realistic five-element array, the enhancement is $\sim 1.8$ times which is significant. \\

\noindent
{\bf Array factor intensity}

\noindent
Let us compare the far-field array factor intensity for the uniform fundamental mode in our SUSY array with a non-uniform fundamental mode which can be obtained by simply setting all the diagonal elements of ${\cal H}_{\rm A}$ to be equal (here we set them equal to $-2t$ with $t=1$). For this array, the (normalized) fundamental mode is given by $a_j=\sqrt{2/(N+1)}\sin[\pi j/(N+1)]$, consequently, the array factor intensity for this non-uniform fundamental mode is 
\begin{widetext}
 \begin{align}
 |{\cal A}(X=0)|^2 &= \frac{2}{N+1} \left|\sum_{j=1}^{N} \sin{\frac{\pi j}{N+1}} \right|^2 \nonumber \\
 &= \frac{1}{2(N+1)} \left|\sum_{j=1}^{N} e^{i\pi j/(N+1)} - {\rm h.c.}\right|^2 \nonumber \\
 &= \frac{1}{2(N+1)} \left|\frac{1-e^{i\pi N/(N+1)}}{1-e^{i\pi/(N+1)}}e^{i\pi/(N+1)} - \frac{1-e^{-i\pi N/(N+1)}}{1-e^{-i\pi/(N+1)}}e^{-i\pi/(N+1)}\right|^2 \nonumber \\
 &= \frac{1}{2(N+1)} \left|\frac{1+e^{-i\pi/(N+1)}}{e^{-i\pi/(N+1)}-1} - \frac{1+e^{i\pi/(N+1)}}{e^{i\pi/(N+1)}-1}\right|^2 \nonumber \\
 &= \frac{1}{2(N+1)} \left|\frac{1+e^{-i\pi/(N+1)}}{1-e^{-i\pi/(N+1)}} + \frac{e^{i\pi/(N+1)}+1}{e^{i\pi/(N+1)}-1}\right|^2 \nonumber \\
 &= \frac{2}{N+1}{\rm cot}^2\left[\frac{\pi}{2(N+1)}\right].
\end{align}
\end{widetext} 

\noindent
{\bf Power distribution inside the laser arrays}

\begin{figure}
 \centering
  \includegraphics[width=\columnwidth]{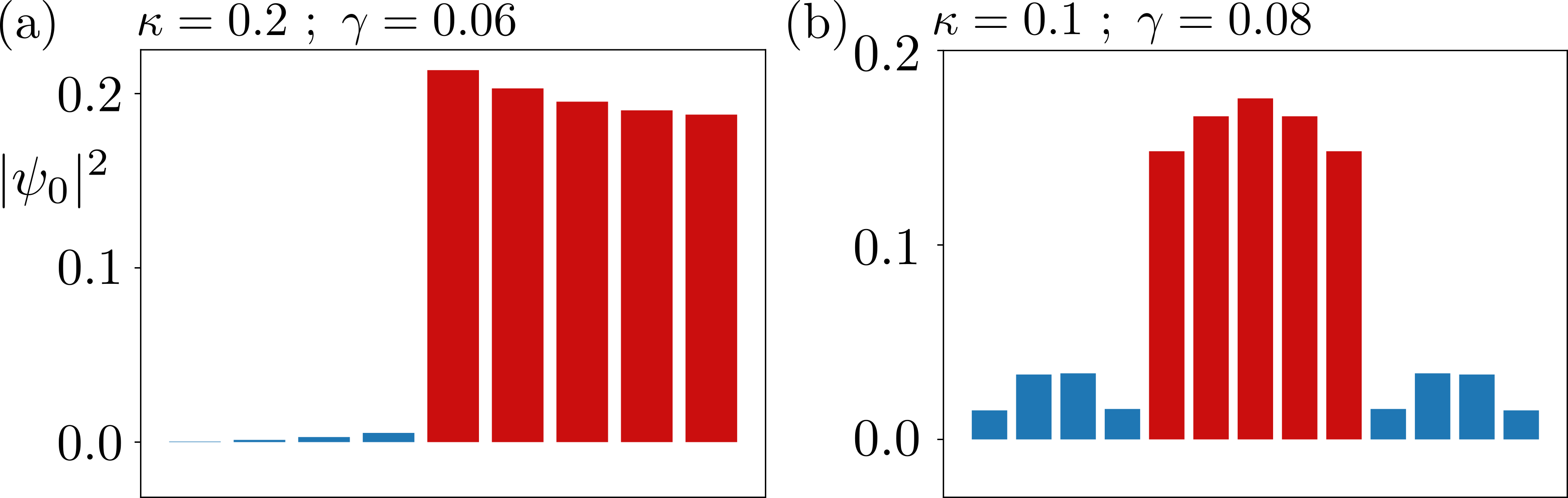}
  \caption{\textbf{Optical power distribution.} Optical power associated with the fundamental modes of both the edge- [in (a)] and surface-emitting [in (b)] SUSY arrays described in the main text. Evidently, most of the optical power is concentrated in the main lasing array (red) with small residual power in the partner arrays (blue). Note that in the case of the edge-emitting laser in panel (a), the power distribution shows a slight asymmetry as compared to the surface-emitting laser in panel (b). This is because, for the edge-emitting laser, we have used only one partner array on one side. The inclusion of a mirror-symmetric partner array on the other side of the laser system would result in a symmetric pattern.}
 \label{Fig_fundamental_mode}
\end{figure}

\noindent
For the proposed SUSY laser to operate properly with far-field enhanced intensity and high beam quality, the optical power must be localized primarily in the main lasing array. Fig.~\ref{Fig_fundamental_mode} confirms that this is indeed the case for both arrays discussed in the main text. These results are also consistent with the recent work on large-area quasi-PT symmetric laser systems described in Ref.~\onlinecite{Ganainy_QPT_LSA_2023}. \\

\noindent
{\bf Data availability:}\\
Data are available upon reasonable request. \\

\noindent
{\bf Code availability:} \\
Codes are available upon reasonable request. \\

\bibliography{ref} 

\vspace{3mm}

\noindent
{\bf Acknowledgments}
R.E. acknowledges support from the Air Force Office of Scientific Research (AFOSR) Multidisciplinary University Research Initiative (MURI) Award on Programmable Systems with non-Hermitian quantum dynamics (Award No. FA9550-21-1-0202), and also from the Alexander von Humboldt Foundation.\\

\noindent
{\bf Author contributions} \\
K.R. and R.E-G. conceived the project and the theoretical formulation. S.D. and M.A. carried out the numerical calculations. All authors contributed to drafting the manuscript. \\

\noindent
{\bf Competing interests} \\
The authors declare no competing interests.

\end{document}